\documentclass[]{spie}  

\usepackage{amsmath,amsfonts,amssymb}
\usepackage{booktabs}
\usepackage{graphicx}
\usepackage{subcaption}
\usepackage[colorlinks=true, allcolors=blue]{hyperref}

\title{Harnessing XGBoost for Robust Biomarker Selection of Obsessive-Compulsive Disorder (OCD) from Adolescent Brain Cognitive Development (ABCD) data}

\author[a]{Xinyu Shen*}
\author[a]{Qimin Zhang}
\author[a]{Huili Zheng}
\author[a]{Weiwei Qi}
\affil[a]{Department of Biostatistics, Columbia University, New York, NY 10032, USA}

\authorinfo{Further author information: (Send correspondence to Xinyu Shen) \\ Xinyu Shen.: E-mail: xs2384@columbia.edu \\ Qimin Zhang: E-mail: qimin.zhang@columbia.edu}

\begin{document} 
\maketitle

\begin{abstract}
This study evaluates the performance of various supervised machine learning models in analyzing highly correlated neural signaling data from the Adolescent Brain Cognitive Development (ABCD) Study, with a focus on predicting obsessive-compulsive disorder scales. We simulated a dataset to mimic the correlation structures commonly found in imaging data and evaluated logistic regression, elastic networks, random forests, and XGBoost on their ability to handle multicollinearity and accurately identify predictive features. Our study aims to guide the selection of appropriate machine learning methods for processing neuroimaging data, highlighting models that best capture underlying signals in high feature correlations and prioritize clinically relevant features associated with Obsessive-Compulsive Disorder (OCD).
\end{abstract}

\keywords{Machine Learning, Feature Selection, Biomarker, Neuroscience}

\section{INTRODUCTION}
\label{sec:intro}  
The Adolescent Brain Cognitive Development (ABCD) Study\cite{casey2018adolescent} represents one of the largest efforts to track brain development and health outcomes in adolescents. Neuroimaging data from this cohort provide a unique opportunity to explore the neural correlates of various cognitive and behavioral conditions, including obsessive-compulsive disorder\cite{boedhoe2018distinct}. 

The rapid growth of the bioinformatics field has led to an increasing reliance on machine learning techniques for the diagnosis and prediction of complex diseases based on their biomarkers\cite{krassowski2022transparent}. However, the high-dimensional nature of biomedical data, with a large number of variables but limited observation data, and multicollinearity, all could pose significant challenges. To address this, researchers have proposed and adopted various classification algorithms for bioinformatics\cite{li2022automated}\cite{li2024leveraging}, including logistic regression, tree-based methods, and deep learning methods, which have emerged as powerful tools for capturing complex patterns across various domains, including image recognition\cite{zhu2024cross}\cite{liu2024image}\cite{shen2024localization}\cite{wang2024research2}, and natural language processing\cite{wang2024research}\cite{liu2024rumor}\cite{zhao2024utilizing}. 

In this study, high correlations between neuroimaging features pose an important challenge and may lead to collinearity in predictive modeling. This study utilized a simulation environment mirroring imaging data from the ABCD study to examine the effectiveness of several supervised machine learning models. Our goal was to identify which models robustly handle the challenge of multicollinearity and accurately predict OCD scales, thereby providing insights into the neurobiological basis of OCD and providing insights for clinical intervention.

\section{Data}
The primary data set for this study comes from the American Adolescent Brain Cognitive Development (ABCD) Study\cite{casey2018adolescent}, the largest long-term study of brain development and child health in the United States. The dataset includes neuroimaging data from a diverse cohort of adolescents, such as structural and functional MRI scans, as well as a wide range of behavioral and health assessments. The clinical endpoint of this study was the Obsessive-compulsive Disorder Scale, which was derived from detailed psychological assessments in the ABCD data and was used to measure the severity and presence of obsessive-compulsive traits among participants.

Patients who have complete obsessive-compulsive disorder scale and corresponding neuroimaging scan data at baseline were included under the selection criteria. After the QC, the sample size should provide enough statistical power for fitting the machine learning models.

Preprocessing of neuroimaging data involves multiple steps to ensure the quality and consistency required for effective machine learning analysis. This includes normalizing imaging data to a standard template, correcting for head motion, and smoothing images to increase signal-to-noise ratio. Additionally, feature extraction was performed to extract meaningful metrics from the raw imaging data, such as cortical thickness, surface area, and subcortical volume, known to be relevant in the context of OCD.

\section{Simulation Method}
\subsection{Simulated Data Generation}

Simulated data allows for the generation of tailored test cases with known ground truth, enabling a more rigorous and comprehensive evaluation of algorithms. The use of simulated data can significantly accelerate the development cycle and improve the overall quality of algorithms by allowing for the testing of edge cases and controlled conditions\cite{mao8633018}\cite{liu2024adaptive}.

The primary challenges encountered in modeling with ABCD data arise from the high degree of multicollinearity, as well as the pronounced imbalance observed in the outcome variable. In order to identify the optimal machine learning model suited to such data characteristics, we generated simulated data that exhibits these attributes.

\begin{figure}[hbt!]
    \centering
    \includegraphics[width=0.75\linewidth]{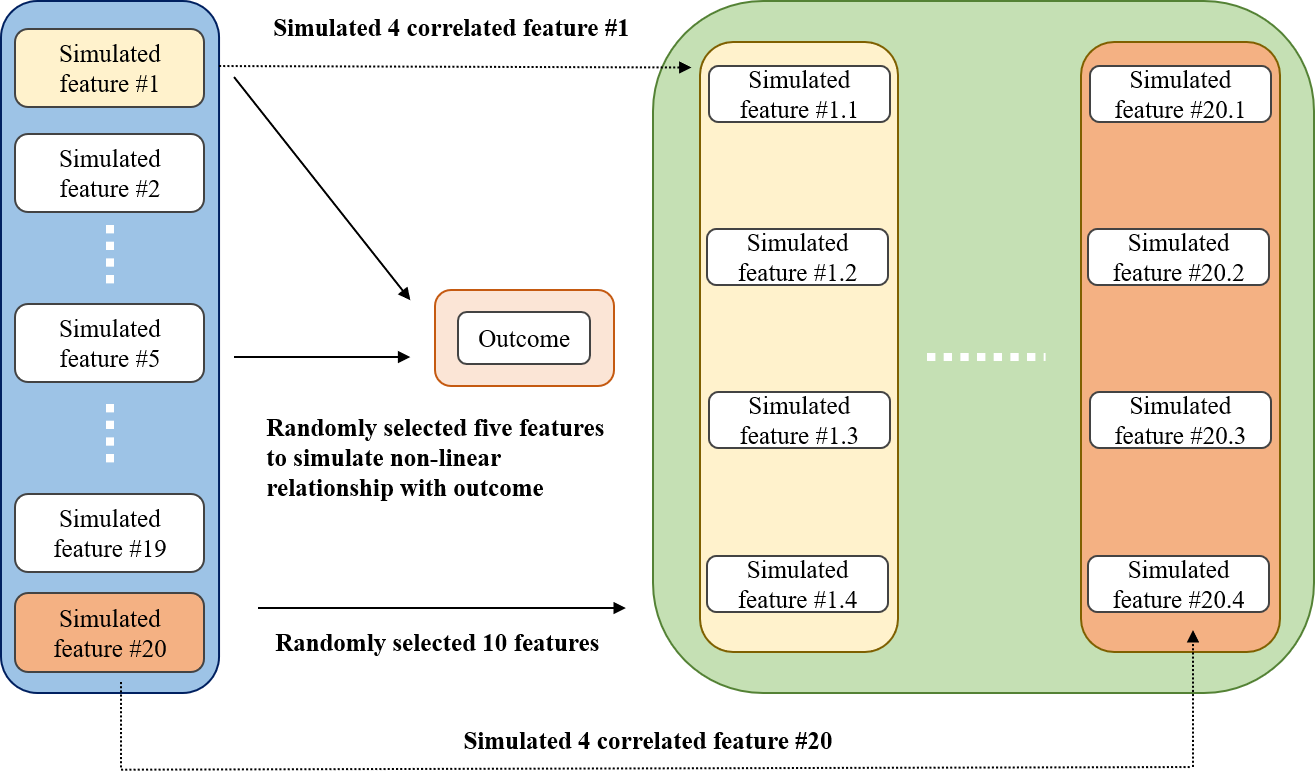}
    \caption{Simulation flow chart. We simulated 10,000 rows with 40 features. Out of 40 features, there are 5 features which have non-linear relationship to the outcome}
    \label{fig:simulation-graph}
\end{figure}

For the simulated dataset, we specify the number of rows to be 10,000 and initialize it with 20 features. It is important to note that these initial 20 features are entirely independent of one another. Subsequently, we randomly select 5 of these features to generate the outcome variable, ensuring that these 5 features are predictive on the outcome:

\begin{equation}
\label{eq:fov}
Y = \sum_{i=1}^{5} \beta_{i1} \cdot x_i^2 + \beta_{i2} \cdot x_i + \epsilon_i
\end{equation}

where $Y$ is the outcome, $x_i$'s are the predictive features, $\beta_{i1}$ and $\beta_{i2}$ are randomly generated coefficients, and $\epsilon_i$ is an error term that follows normal distribution $\mathcal{N}(0, 1)$. We introduced this non-linear relationship to simulate real-world conditions observed in ABCD data.

Subsequently, within the set of 20 features, we randomly select 10 features to create additional correlated features. For each of the 10 selected features, we generate 4 new features that exhibit linear correlation. As a result of this procedure, the dataset encompasses a total of 50 features. In the end, we dichotomized $Y$ to be binary.

\subsection{Candidate Machine Learning Models}
We primarily evaluated the performance of the simulated data using several mainstream machine learning models, including logistic regression, elastic-net, random forest, and XGBoost\cite{xgboost}. The dataset was partitioned into an 80\% training set and a 20\% test set, and the performance of each model was assessed using the Area Under the Curve (AUC) score.

\subsection{Simulation Results and Analysis}

\begin{figure}[htbp]
    \centering
    \begin{subfigure}[b]{0.45\textwidth}
        \includegraphics[width=\textwidth]{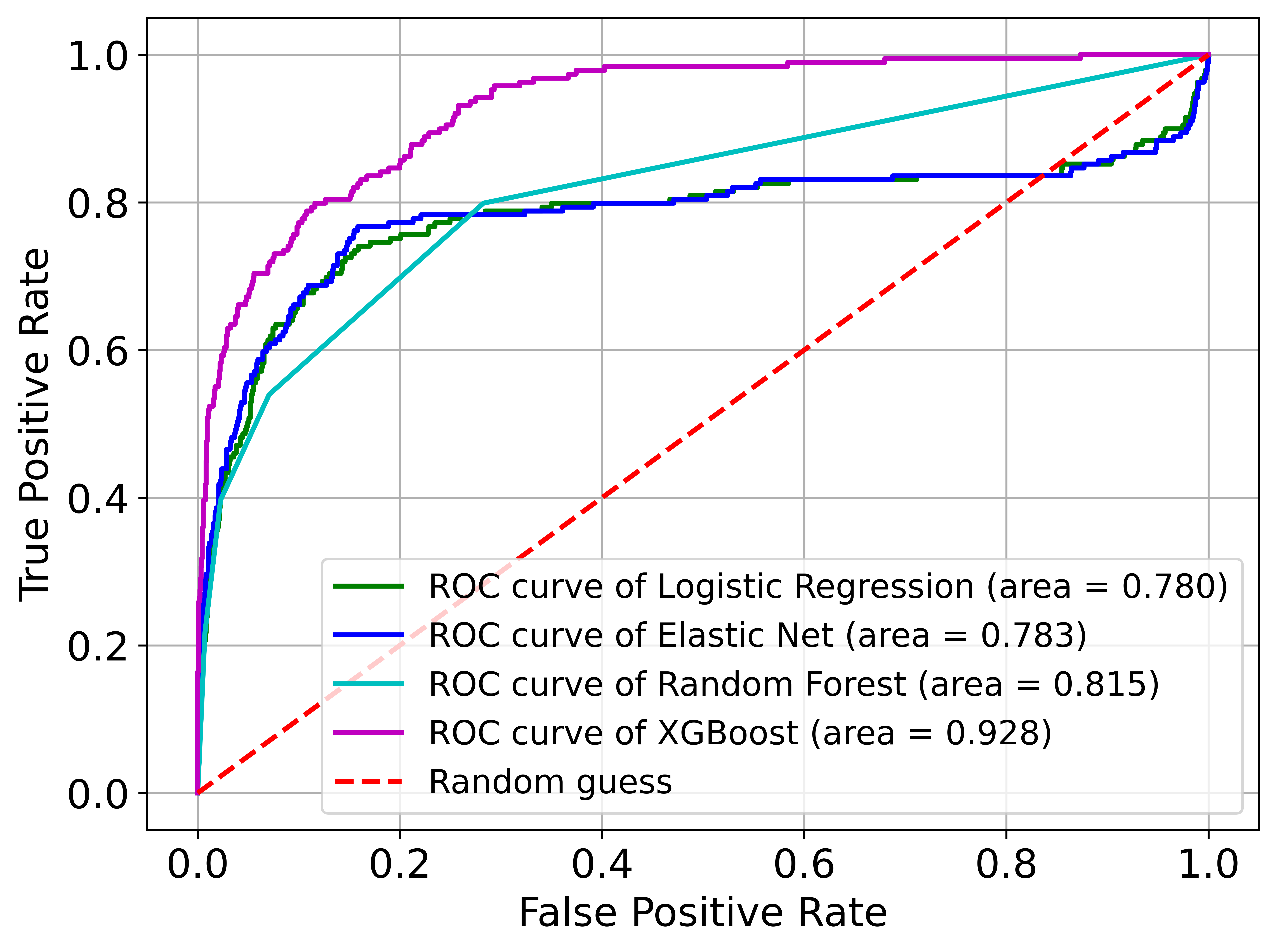}
        \caption{ROC curves of candidate machine learning models}
        \label{fig:roc}
    \end{subfigure}
    \hfill
    \begin{subfigure}[b]{0.45\textwidth}
        \includegraphics[width=\textwidth]{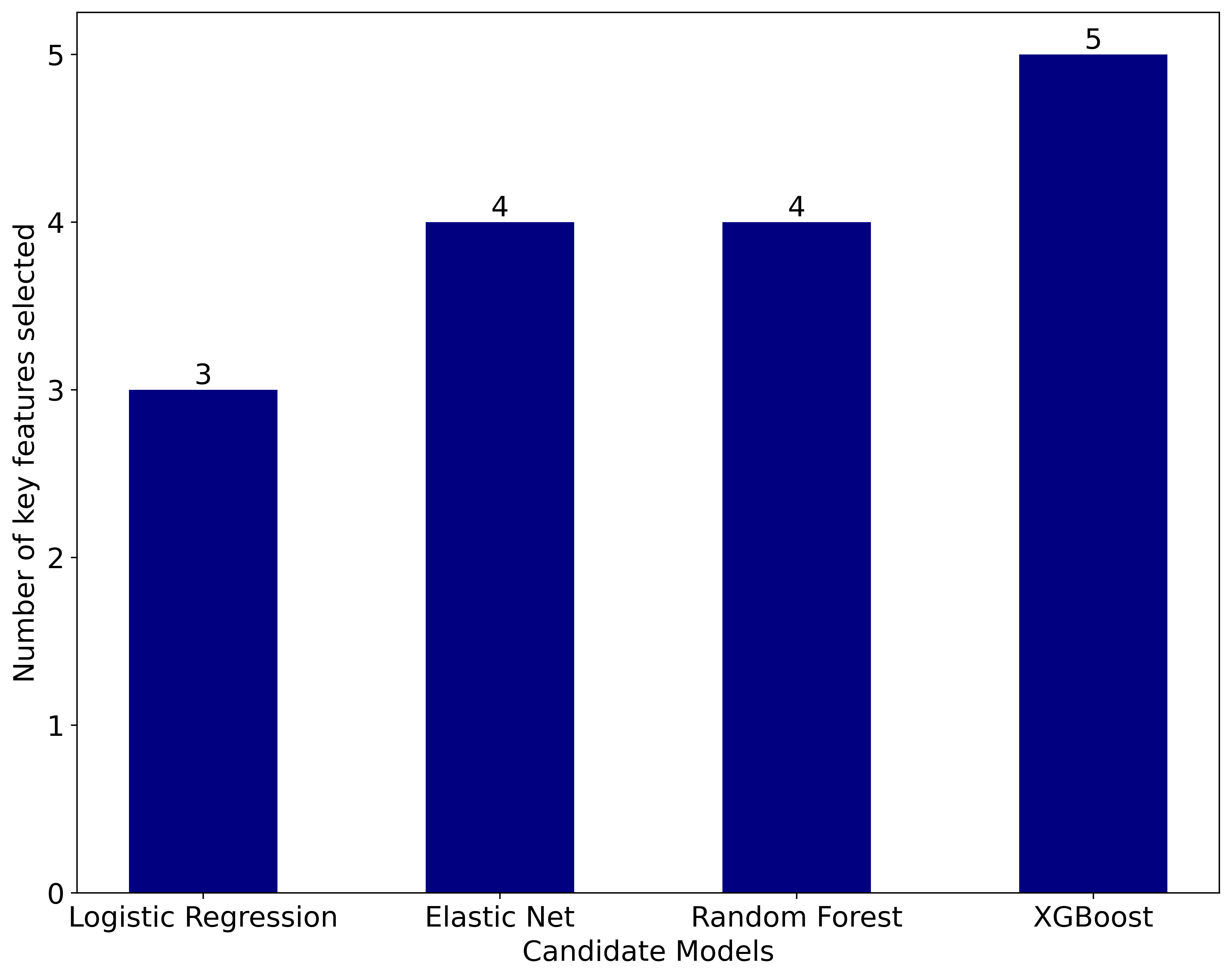}
        \caption{Number of key features selected among top 5 features of candidate models}
        \label{fig:nfs}
    \end{subfigure}
    \caption{Simulation results}
    \label{fig:sr}
\end{figure}

As shown in Fig \ref{fig:roc}, among the 4 candidate machine learning models, XGBoost had the performance in terms of AUC score, with AUC score of 0.928. In addition, as shown in Fig \ref{fig:nfs}, XGBoost successfully captured all 5 predictive features as its top 5 features, while Logistic Regression, Elastic-net and Random Forest only captured 3, 4 and 4 respectively. This demonstrates the ability of XGBoost to capture the most predictive features.

We can also notice the the 2 linear methods (logistic regression and Elastic-Net) underperformed the other 2 non-linear ones. That's because if the underlying relationship between the input features and the target variable is inherently non-linear, a linear model may not be able to capture the complex patterns in the data. Non-linear models like random forests are more flexible and can better fit complex, non-linear functions\cite{Elgart2022}. In addition, linear models assume that the effect of each feature on the target variable is independent. However, in this study, there are feature interactions that a linear model cannot capture. Non-linear models like random forests can automatically detect and model these complex feature interactions\cite{Elgart2022}.

\section{XGBoost}
We then determined to use XGBoost on the ABCD data to identify biomarkers that contribute to OCD. The objective function\cite{xgboost} is defined as:
\begin{equation}
\text{Objective}(T) = \sum_{i=1}^{n} l(y_i, \hat{y}i) + \sum{f \in T} \Omega(f)
\end{equation}
where $T$ represents the ensemble of decision trees, $l(y, \hat{y})$ is a differentiable convex loss function that measures the difference between the true output $y$ and the predicted output $\hat{y}$, $y_i$ is the true output for instance $i$, $\hat{y}_i$ is the predicted output for instance $i$, and $\Omega(f)$ is the regularization term applied to each tree $f$ in the ensemble $T$.

XGBoost learns the target function in an additive manner, creating an iterative ensemble of decision trees (weak learners) that gradually minimizes the objective function \cite{xgboost}. In each iteration, a new tree is added to the ensemble, and the objective function is optimized. This can be formalized as:
\begin{equation}
F_m(x) = F_{m-1}(x) + f_m(x)
\end{equation}
where $F_m(x)$ is the prediction after adding $m$ trees, $F_{m-1}(x)$ is the prediction after adding $m-1$ trees, and $f_m(x)$ is the new tree added in the $m$-th iteration.
\begin{figure}
    \centering
    \includegraphics[width=0.75\linewidth]{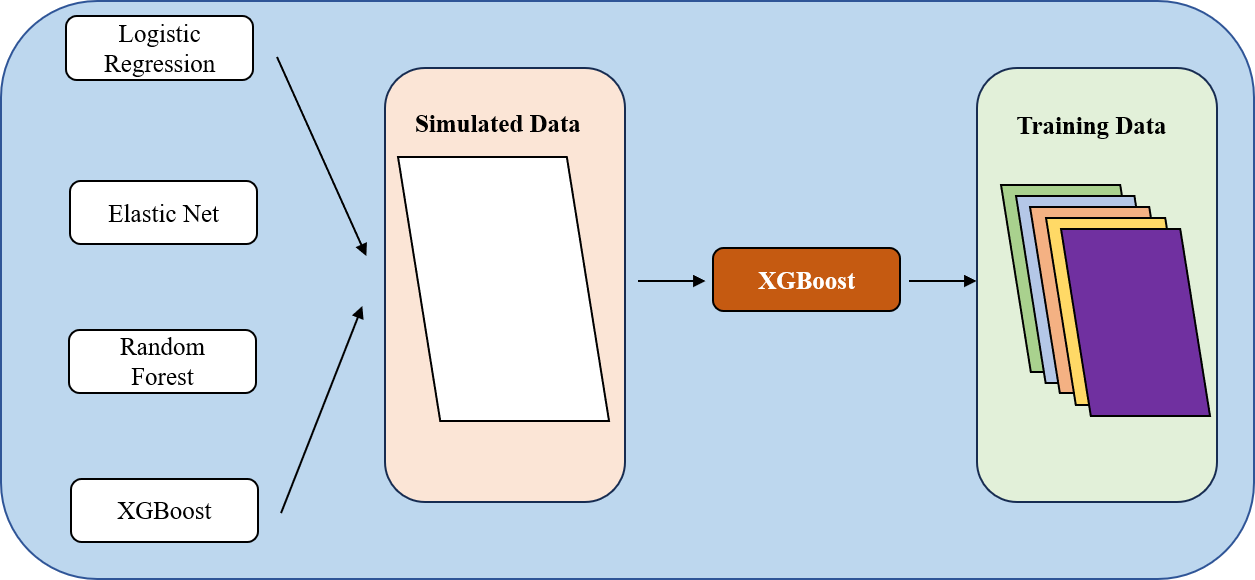}
    \caption{XGBoost as the best classifier: Applied XGBoost into training data}
    \label{fig:ml-process}
\end{figure}
XGBoost demonstrates cache-awareness, mitigating overfitting by managing model complexity and integrating built-in regularization techniques. It adeptly manages sparse data and extends its capabilities to utilize disk space for large datasets, enabling out-of-core computing. This maximizes system resources, enhances computational efficiency, and ultimately improves prediction performance\cite{li2024utilizinglgbm}. In our case, we expect XGBoost to capture complex patterns in ABCD data to find biomarkers that are assoicated with OCD.

\section{Results}

   \begin{figure} [ht]
   \begin{center}
   \begin{tabular}{c} 
   \includegraphics[height=5cm]{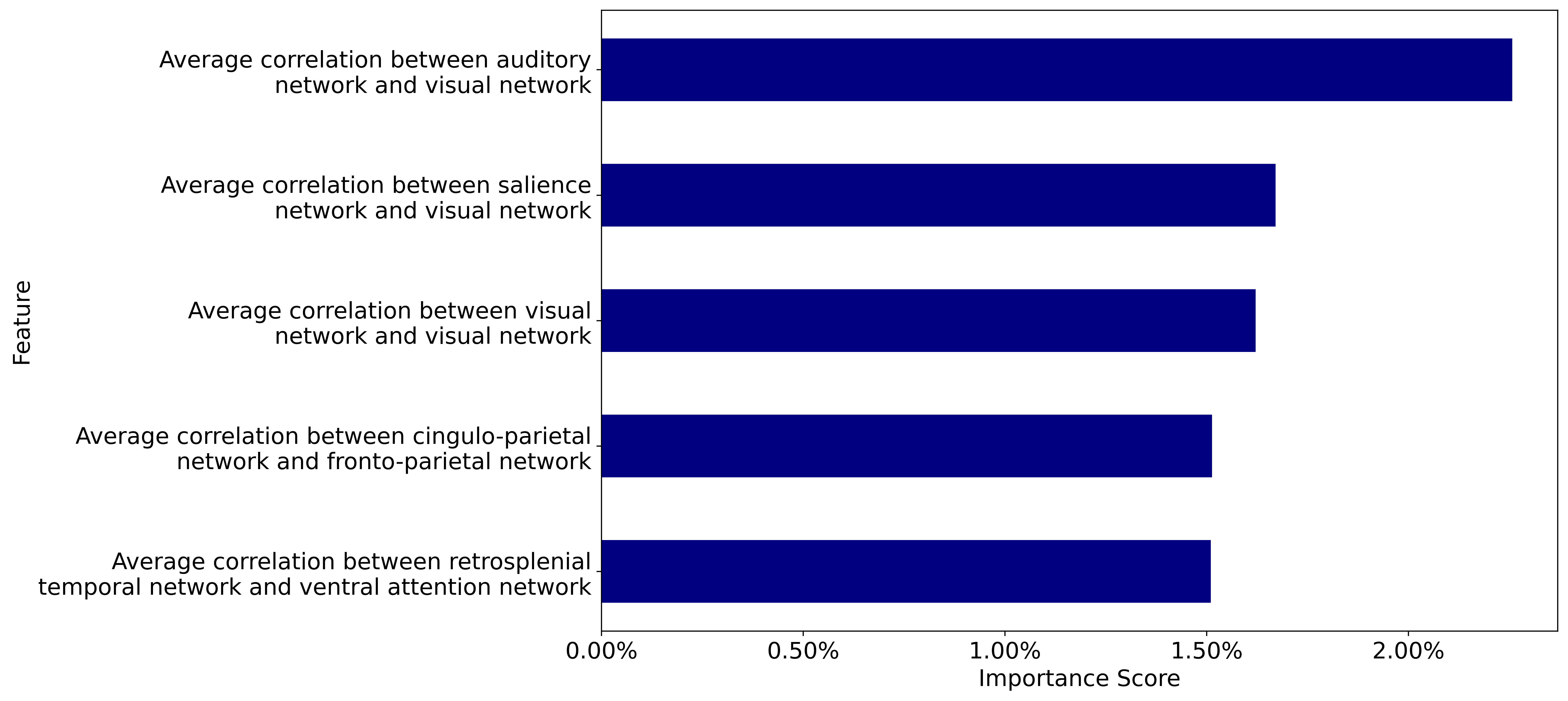}
	\end{tabular}
	\end{center}
   \caption[example] 
    { \label{fig:feature-importance} 
    Feature importance from XGBoost on ABCD data}
   \end{figure}

As shown in the Matplotlib\cite{Weng2024cyber}-visualized feature importance fig \ref{fig:feature-importance}, the top 3 biomarkers selected from real-world ABCD dataset by XGBoost are: Average correlation between auditory network and visual network, Average correlation between salience network and visual network, Average correlation between visual network and visual network, Average correlation between cingulo-parietal network and fronto-parietal network, and Average correlation between retrosplenial temporal network and ventral attention network.

The visual network in the brain is a complex, distributed system responsible for processing various aspects of visual information, including color, shape, motion, and object recognition\cite{Brissenden2018}. The auditory network is a complex, distributed system in the brain responsible for cortical auditory function and likely plays a crucial role in cognitive and language functions\cite{Tobyne2017}. The salience network is a large-scale brain network detecting and filtering salient, biologically and cognitively relevant stimuli, recruiting and modulating the activity of other relevant brain networks, such as the default mode network and central executive network, to guide flexible behavior, and integrating sensory, emotional, and cognitive information to support complex functions like communication, social behavior, and self-awareness\cite{seeley2007dissociable}. 

We can see that the top 3 biomarkers selected are all correlation between visual network and another network handling different types of stimulus, which reflects the subjects with clinical significant OCD are with enhanced functional integration and communication between these systems\cite{Jao2021}. It could also be a compensatory mechanism, where the brain increases connectivity to enhance processing of sensory information in the face of deficits or impairments\cite{Jao2021}. In addition, it could emerge as a result of extensive training or expertise, reflecting adaptive changes in brain organization\cite{Gong2016}.

\section{Conclusions}
In our investigation of various mainstream machine learning models, XGBoost emerged as the top performer across multiple challenging scenarios. Specifically, it demonstrated superior performance in handling multicollinearity, effectively managing predictors with non-linear relationships with the target variable, and addressing imbalanced data. XGBoost outperformed other mainstream classification method in the designed simulated data, which successfully captured all true signals hidden by the correlated features and thus have the highest AUC score. These findings underscore the versatility and robustness of XGBoost in navigating complex data structures and highlight its potential as a preferred choice for addressing real-world problems characterized by such challenges. The high performance of XGBoost from simulation experiment confirmed the robustness of applying it in a highly correlated settings and thus give the confidence of applying it into real-world data.

In our experimentation with real-world ABCD data, XGBoost selected the top biomarkers, which predominantly represent the visual network and others responsible for processing different types of stimuli. This selection suggests that individuals with clinically significant OCD exhibit heightened functional integration and communication among these networks. This phenomenon may indicate a compensatory mechanism where increased connectivity enhances sensory information processing in response to deficits or impairments. Additionally, it could signify adaptive changes in brain organization due to extensive training or expertise. Such findings imply a potential augmentation in situational awareness and responsiveness among these individuals, shedding light on the neurocognitive mechanisms underlying OCD and presenting avenues for further exploration in clinical and cognitive neuroscience.

\bibliography{report} 

\begin{thebibliography}{10}

\bibitem{casey2018adolescent}
Casey, B., Cannonier, T., Conley, M., Cohen, A., Barch, D., Heitzeg, M., Soules, M., Teslovich, T., Dellarco, D., Garavan, H., et~al., ``The adolescent brain cognitive development (abcd) study: imaging acquisition across 21 sites,'' {\em Developmental cognitive neuroscience}~{\bf 32},  43--54 (2018).

\bibitem{boedhoe2018distinct}
Boedhoe, P., Schmaal, L., Abe, Y., Alonso, P., Ameis, S., Anticevic, A., Calvo, R., Carpenter, D., Cho, K., Dallaspezia, S., et~al., ``Distinct subcortical volume alterations in pediatric and adult ocd: a worldwide meta-and mega-analysis,'' {\em American Journal of Psychiatry}~{\bf 175}(5),  435--444 (2018).

\bibitem{krassowski2022transparent}
Krassowski, M., Paczkowska, M., Cullion, K., Huang, T., Rosen, G., and Przytycka, T.~M., ``Transparent exploration of machine learning for biomarker discovery,'' {\em Journal of Proteome Research}~{\bf 21}(12),  2524--2534 (2022).

\bibitem{li2022automated}
Li, S., Mo, Y., and Li, Z., ``Automated pneumonia detection in chest x-ray images using deep learning model,'' {\em Innovations in Applied Engineering and Technology} ,  1--6 (2022).

\bibitem{li2024leveraging}
Li, S., Qu, H., Dong, X., Dang, B., Zang, H., and Gong, Y., ``Leveraging deep learning and xception architecture for high-accuracy mri classification in alzheimer diagnosis,'' {\em arXiv preprint arXiv:2403.16212}  (2024).

\bibitem{zhu2024cross}
Zhu, A. et~al., ``Cross-task multi-branch vision transformer for facial expression and mask wearing classification,'' {\em arXiv preprint arXiv:2404.14606}  (2024).

\bibitem{liu2024image}
Liu, T., Cai, Q., Xu, C., Hong, B., Xiong, J., Qiao, Y., and Yang, T., ``Image captioning in news report scenario,'' {\em Academic Journal of Science and Technology}~{\bf 10}(1),  284--289 (2024).

\bibitem{shen2024localization}
Shen, Y., Liu, H., Liu, X., Zhou, W., Zhou, C., and Chen, Y., ``Localization through particle filter powered neural network estimated monocular camera poses,'' {\em arXiv preprint arXiv:2404.17685}  (2024).

\bibitem{wang2024research2}
Wang, J., Li, X., Jin, Y., Zhong, Y., Zhang, K., and Zhou, C., ``Research on image recognition technology based on multimodal deep learning,'' (2024).

\bibitem{wang2024research}
Wang, J. et~al., ``Research on emotionally intelligent dialogue generation based on automatic dialogue system,'' {\em arXiv preprint arXiv:2404.11447}  (2024).

\bibitem{liu2024rumor}
Liu, T., Cai, Q., Xu, C., Hong, B., Ni, F., Qiao, Y., and Yang, T., ``Rumor detection with a novel graph neural network approach,'' {\em Academic Journal of Science and Technology}~{\bf 10}(1),  305--310 (2024).

\bibitem{zhao2024utilizing}
Zhao, Y. and Gao, H., ``Utilizing large language models for information extraction from real estate transactions,'' {\em arXiv preprint arXiv:2404.18043}  (2024).

\bibitem{mao8633018}
Mao, C., Wang, J., and Tao, L., ``A new sidelobe suppression algorithm for sar images with an arbituary doppler centroid,'' in [{\em 2018 11th International Congress on Image and Signal Processing, BioMedical Engineering and Informatics (CISP-BMEI)}{\nolinebreak\hspace{0.1em}]},   1--5 (2018).

\bibitem{liu2024adaptive}
Liu, H., Shen, Y., Zhou, W., Zou, Y., Zhou, C., and He, S., ``Adaptive speed planning for unmanned vehicle based on deep reinforcement learning,'' {\em arXiv preprint arXiv:2404.17379}  (2024).

\bibitem{xgboost}
Chen, T. and Guestrin, C., ``Xgboost: A scalable tree boosting system.'' Software (2016).
\newblock \url{https://github.com/dmlc/xgboost}.

\bibitem{Elgart2022}
Elgart, M., Lyons, G., Romero-Brufau, S., Kurniansyah, N., Brody, J.~A., Guo, X., Lin, H.~J., Raffield, L., Gao, Y., Chen, H., et~al., ``Non-linear machine learning models incorporating snps and prs outperform linear models for complex human phenotypes,'' {\em Nature Communications}~{\bf 13}(1),  1--12 (2022).

\bibitem{li2024utilizinglgbm}
Li, S., Dong, X., Ma, D., Dang, B., Zang, H., and Gong, Y., ``Utilizing the lightgbm algorithm for operator user credit assessment research,'' {\em arXiv preprint arXiv:2403.14483}  (2024).

\bibitem{Weng2024cyber}
Weng, Y. and Wu, J., ``Fortifying the global data fortress: a multidimensional examination of cyber security indexes and data protection measures across 193 nations,'' {\em International Journal of Frontiers in Engineering Technology}~{\bf 6}(2) (2024).

\bibitem{Brissenden2018}
Brissenden, J.~A. and Somers, D.~C., ``Redefining the functional organization of the visual system,'' {\em Annual Review of Vision Science}~{\bf 4},  355--378 (2018).

\bibitem{Tobyne2017}
Tobyne, S.~M., Langley, J., Binney, R.~J., Rorden, C., and Olson, I.~R., ``Resting-state functional connectivity in an auditory network differs between individuals with and without bothersome tinnitus,'' {\em NeuroImage: Clinical}~{\bf 17},  509--515 (2018).

\bibitem{seeley2007dissociable}
Seeley, W.~W., Menon, V., Schatzberg, A.~F., Keller, J., Glover, G.~H., Kenna, H., Reiss, A.~L., and Greicius, M.~D., ``Dissociable intrinsic connectivity networks for salience processing and executive control,'' {\em Journal of Neuroscience}~{\bf 27}(9),  2349--2356 (2007).

\bibitem{Jao2021}
Jao~Keehn, R., Pueschel, E., Gao, Y., Jahedi, A., Alemu, K., Carper, R., Fishman, I., and Müller, R., ``Underconnectivity between visual and salience networks and links with sensory abnormalities in autism spectrum disorders,'' {\em Journal of the American Academy of Child \& Adolescent Psychiatry}~{\bf 60}(2),  274--285 (2021).

\bibitem{Gong2016}
Gong, D., He, H., Ma, W., Liu, D., Huang, M., Dong, L., Gong, J., Li, J., Luo, C., and Yao, D., ``Functional integration between salience and central executive networks: A role for action video game experience,'' {\em Neural Plasticity}~{\bf 2016},  9803165 (2016).

\end{thebibliography}
\bibliographystyle{spiebib} 

\end{document}